\documentclass[12pt,letterpaper]{article}
\usepackage[usenames, dvipsnames]{xcolor}
\usepackage{jcapmod}

\usepackage{tocloft}
\usepackage[]{todonotes}
\usepackage{bbold}
\usepackage{graphicx}
\usepackage{verbatim}
\usepackage[mathscr]{euscript}
\usepackage{slashed}
\usepackage{mathdots}
\usepackage{caption}

\usepackage{chngcntr}
\counterwithout{equation}{section}

\usepackage[labelformat=simple]{subcaption}

\usepackage{enumerate}
\usepackage{bbm}
\usepackage{psfrag}
\usepackage{subfiles}
\usepackage{psfrag}
\usepackage{relsize}
\usepackage[T1]{fontenc}
\usepackage[utf8]{inputenc}
\usepackage{pgfplots}
\usepackage{tikzscale}

\usepackage{bbm} 					
\usepackage{slashed} 				
\usepackage{graphicx}				
\usepackage{subcaption}			
\usepackage{psfrag}				
\usepackage{tensor}				
\usepackage{fouridx}				
\usepackage{bm}					
\usepackage{mdframed}				
\usepackage{multirow}				
\usepackage{soul}					
\usepackage{bbold}				
\usepackage{multicol}				

\usepackage{amsmath}
\usepackage{amssymb}
\usepackage{amsthm}

\usepackage{mathtools}

\usepackage{feynmf}
\usepackage{marvosym}

\usepackage{import}

\newtheoremstyle{named}{1pt}{1pt}{\itshape}{}{\bfseries}{.}{.5em}{#3}
\theoremstyle{named}
\newtheorem*{namedconjecture}{Conjecture}

\newcommand{\e}{{\mathrm{e}}}

\newcommand{\cM}{\mathcal{M}}

\newcommand{\rmd}{\textrm{d}}
\def\be{\begin{equation}}
\def\ee{\end{equation}}
\def\bea{\begin{eqnarray}}
\def\eea{\end{eqnarray}}

\ProvideTextCommandDefault{\Dbar}{%
\leavevmode\lower.5ex\rlap{\hskip-.07em\accent"16}D%
}

\usepackage{setspace}
\usepackage[margin=1.0in]{geometry}

\usepackage{titlesec}
\titlespacing*{\section} {0pt}{0.75ex}{0.74ex}

\begin{document}
	\newcommand{\main}{.}
\begin{titlepage}

\setcounter{page}{1} \baselineskip=12pt \thispagestyle{empty}

\bigskip\

\vspace{1cm}
\begin{center}
{\fontsize{19}{24} \bfseries A Symmetry-centric Perspective on the Geometry \\\vspace{0.3cm} of the String Landscape and the Swampland}

 \end{center}
\vspace{1cm}

\begin{center}
\scalebox{0.95}[0.95]{{\fontsize{14}{30}\selectfont  Tom Rudelius$^\dagger$}}\\\vspace{0.25 cm}
{\small \color{gray} \texttt{thomas.w.rudelius@durham.ac.uk}}\\
\end{center}
\begin{center}
\vspace{0.25 cm}
\textsl{$^{\dagger}$Department of Mathematical Sciences, Durham University, \\ \vspace{0.1 cm}
Upper Mountjoy Campus, Stockton Rd, Durham DH1 3LE UK}\\

\vspace{0.25cm}

\end{center}

\vspace{.9cm}
\noindent 
\onehalfspacing
As famously observed by Ooguri and Vafa nearly twenty years ago, scalar field moduli spaces in quantum gravity appear to exhibit various universal features. For instance, they seem to be infinite in diameter, have trivial fundamental group, and feature towers of massive particles that become light in their asymptotic limits.
In this essay, we explain how these features can be reformulated in more modern language using generalized notions of global symmetries. Such symmetries are ubiquitous in non-gravitational quantum field theories, but it is widely believed that they must be either gauged or broken in quantum gravity. In what follows, we will see that the observations of Ooguri and Vafa can be understood as consequences of such gauging or breaking.

 \vspace{.8cm}
 \noindent
\emph{Essay written for the Gravity Research Foundation 2024 Awards for Essays on Gravitation.}

 \vspace{.9cm}

\bigskip
\noindent\today

\end{titlepage}
\setcounter{tocdepth}{2}

\section*{Introduction}\label{INTRO}

Scalar fields play a vital role in the physics of our universe. The most famous example of this is the Higgs boson, but scalar fields also feature prominently in models of dark matter, dark energy, and inflation, and in resolving the strong-CP problem. Scalar fields also lie at the heart of naturalness puzzles such as the hierarchy problem and the cosmological constant problem, and thus they may offer a unique window into the inner workings of the ever-mysterious theory of quantum gravity.

Nearly twenty years ago, the quest to understand the effects of quantum gravity on scalar field theories led Ooguri and Vafa to publish their seminal work, ``On the Geometry of the String Landscape and the Swampland'' \cite{Ooguri:2006in}, which proposed and provided evidence for five conjectures about scalar fields in quantum gravity. A counterexample to one of the conjectures was discovered shortly thereafter \cite{trenner2010asymptotic}, but the other four conjectures remain conjectures: none of them have been proven, yet no counterexamples have been given, either.

At present, most of the evidence for the Ooguri-Vafa conjectures comes from string theory, which offers a vast landscape of UV-complete scalar field theories coupled to Einstein gravity. One issue with this line of evidence, however, is that most of our knowledge of scalar fields in string theory comes from studies of exactly massless scalar fields, also known as moduli, in theories with unbroken supersymmetry. As a result, it is unclear whether the apparently universal features of these theories will persist once supersymmetry is broken, as it is in our universe.

In this essay, we explain how these conjectures can be reformulated in the language of generalized global symmetries. Such symmetries make sense even in non-supersymmetric quantum field theories, so these reformulations help lend credence to the Ooguri-Vafa conjectures beyond the supersymmetric lamppost and, hopefully, will ultimately lead us toward their proofs.

In a first course on quantum field theory, students learn about ``ordinary'' global symmetries, which transform local operators by elements of a group $G$ in such a way as to keep the Lagrangian invariant. In recent years, however, it has become clear that such symmetries are just an example of a much broader zoo of generalized notions of symmetry: high-form symmetries \cite{Gaiotto:2014kfa} involve symmetry transformations that act not on local operators of the theory, but rather on extended operators (line operators, surface operators, etc.). Higher-group symmetries \cite{Sharpe:2015mja, Cordova:2018cvg, Benini:2018reh} involve two or more higher-form symmetries that intertwine into a larger structure. Non-invertible symmetries involve transformations that are labeled not by group elements, but by elements of a more general category. Many of the features of ordinary symmetries---charged operators, symmetry generators, anomalies, gauging, and so on---carry over to these generalized notions of symmetry as well.

It is widely believed that all of these different types of global symmetries are forbidden in a consistent theory of quantum gravity \cite{Hawking:1974sw, Zeldovich:1976vq, Zeldovich:1977be, banks:1988yz, kallosh:1995hi, Banks:2010zn, Harlow:2018tng, Harlow:2020bee, Chen:2020ojn, Yonekura:2020ino}. However, such symmetries appear readily in quantum field theories, which means that any symmetry must be either broken or gauged upon coupling it to gravity. In what follows, we will discuss each of the Ooguri-Vafa conjectures in turn, and we will see how they may be viewed as consequences of gauging or breaking generalized global symmetries.

\section*{Conjecture 0 and lower-form symmetries}\label{sec:C0}

 The first conjecture of \cite{Ooguri:2006in}, entitled ``Conjecture 0,'' reads as follows:
 
 \vspace{.2cm}
\begin{namedconjecture}[Conjecture 0]
Let $\cM$ be the moduli space of a quantum gravity theory in $d \geq 4$. $\cM$ is parameterized by inequivalent expectation values of massless scalar
fields.
 \end{namedconjecture}
  \vspace{.2cm}

 \noindent
 In other words, Conjecture 0 forbids the existence of (continuous) free parameters in quantum gravity. Coupling constants are determined by vacuum expectation values of scalar fields; they cannot be tuned to generate distinct theories of quantum gravity the way the parameters of the standard model can be tuned to generate different quantum field theories.
 
 Conjecture 0 predates even \cite{Ooguri:2006in} (hence why it was called Conjecture 0 rather than Conjecture 1), but recent progress permits an alternative understanding of Conjecture 0 in terms of generalized global symmetries.
 
 A $p$-form global symmetry in a quantum field theory is a global symmetry for which the charged operators are $p$-dimensional objects. For instance, Wilson line operators may be charged under a 1-form global symmetry, while local operators may be charged under a 0-form global symmetry. Such a symmetry is labeled by a group $G$, and when $G$ is continuous, the theory may be coupled to a background $(p+1)$-form gauge field. Gauging the symmetry then amounts to promoting this background gauge field to a dynamical gauge field by adding a kinetic term for the field to the Lagrangian and integrating over it in the path integral.
 
With this, parameters in quantum field theory can be viewed as $(-1)$-form global symmetries \cite{Cordova:2019jnf, Heidenreich:2020pkc, Vandermeulen:2022edk}: they may be coupled to background scalar fields, and varying these background fields leads to different values of the parameters. More precisely, a compact, circle-valued background scalar field leads to a $U(1)$ $(-1)$-form global symmetry, while a noncompact scalar field leads to an $\mathbb{R}$ $(-1)$-form global symmetry. In this language, Conjecture 0 amounts to the statement that all continuous $(-1)$-form symmetries in quantum gravity must be gauge symmetries.

Conjecture 0 can thus be understood as a special case of the more general conjecture that all symmetries in quantum gravity must be either gauged or broken. At present, however, the arguments against global symmetries in quantum gravity deal primarily with ordinary global symmetries and, to a lesser extent, higher-form global symmetries \cite{Harlow:2018tng}. Extending these arguments to the case of $(-1)$-form symmetries remains a task for future study.\footnote{See however \cite{Heckman:2024oot} for progress in this direction.}

\section*{Conjecture 4 and non-invertible higher-form symmetries}

 \vspace{.2cm}
\begin{namedconjecture}[Conjecture 4]
There is no non-trivial 1-cycle with minimum length within a given homotopy class in $\cM$.
\end{namedconjecture}
 \vspace{.2cm}
 
\noindent
In \cite{Ooguri:2006in}, Ooguri and Vafa argued that a nontrivial fundamental group $\pi_1(\cM)$ would lead to global symmetry upon compactifying to 1+1 dimensions. If global symmetries are not allowed in 2d theories of quantum gravity, then this scenario should not be allowed, either.

From a modern perspective, we can recast this argument from Ooguri and Vafa in terms of $(d-2)$-form symmetries in $d$ dimensions, removing the need to compactify to 2d. Suppose that our theory features an axion $\theta$ with periodic identification $\theta \sim \theta + 2\pi$ and action
\begin{equation}
S = \int \rmd^d x \sqrt{-\mathfrak{g}_d } \left[ - \frac{f^2}{2} (\partial_\mu \theta)^2 \right]\,,
\label{action1}
\end{equation}
where $f$ is a constant known as the axion decay constant. The periodicity of $\theta$ implies that the moduli space $\cM$ is isomorphic to $S^1$, which has $\pi_1(S^1) \cong \mathbb{Z}$, and the fact that $f$ is constant means that there is no way to contract a nontrivial loop in $\pi_1(\cM)$ to a point, in violation of Conjecture 4.

Relatedly, this theory features a family of codimension-2 surface operators, each labeled by an integer $n$, such that $\theta$ winds by $2 \pi n$ as one winds around the surface operator in the transverse space. These operators are charged under a $(d-2)$-form symmetry \cite{Gaiotto:2014kfa}. Indeed, adopting the perspective of the previous section, we may think of the axion as a 0-form gauge field for a $(-1)$-form gauge symmetry, and we may think of these surface operators as 't Hooft surfaces for this gauge field, which are charged under a magnetic $(d-2)$-form symmetry with current $(\star j)_\mu = \partial_\mu \theta$.

More generally, a sigma model with target space $\mathcal{M}$ will have codimension-2 surface operators labeled by conjugacy classes of $\pi_1(\cM)$ \cite{Hsin:2022heo}. These operators are charged under a $(d-2)$-form symmetry whose elements are labeled by representations of $\pi_1(\cM)$. If $\pi_1(\cM)$ is abelian, then the symmetry group is simply the Pontryagin dual of $\pi_1(\cM)$, and it is an abelian group. If $\pi_1(\cM)$ is non-abelian, however, then $(d-2)$-form symmetry in question is known as a non-invertible symmetry, as its elements do not fuse according to a simple group multiplication law ($g * h = gh$), but rather according to the tensor product of representations of $\pi_1(\cM)$ ($\rho_i \times \rho_j = \sum_k C_{ij}^k \rho_k$).

To break this symmetry, we must (a) promote the decay constant $f$ to a dynamical field (as also required by Conjecture 0 above), and (b) demand that this decay constant vanishes at some point in the closure of field space, $\overline\cM$. The simplest (and most common) way to do this is to complete the axion into some version of the abelian Higgs model \cite{Kim:1979if, Shifman:1979if, Zhitnitsky:1980tq, Dine:1981rt}:
\be
S = \int \rmd^d x \sqrt{-\mathfrak{g}_d} \left[  - \frac{1}{2} (\partial_\mu \phi^\dagger)  (\partial^\mu \phi) - \frac{\lambda}{4}  \left(|\phi|^2 - \frac{\mu^2}{\lambda} \right)^2 \right]\,.
\ee
At low energies, $\theta \equiv \arg(\phi)$ behaves like an axion with decay constant $f =\langle |\phi| \rangle = \mu / \sqrt{\lambda}$, and there is an emergent $(d-2)$-form symmetry. However, at higher energies, this symmetry is broken, as winding number is not a conserved quantity at the origin $\phi = 0$. In this model, the point $\phi = 0$ lies at finite distance and is part of the moduli space, so $\pi_1(\cM)$ is trivial, there is no $(d-2)$-form symmetry, and Conjecture 4 is satisfied.

However, within quantum gravity there exists another type of axion, which does not arise as the angular mode of a complex scalar field. Such an axion is called a ``fundamental axion'' \cite{Reece:2018zvv}, and its decay parameter $f$ vanishes only at an infinite-distance boundary of the moduli space.. As a result, winding a fundamental axion around its fundamental domain yields a nontrivial element of $\pi_1(\cM)$, and breaking the associated $(d-2)$-symmetry requires the existence of dynamical, finite-energy axion vortices, which (like the aforementioned surface operators) induce monodromy of the axion field as one winds around their core \cite{Rudelius:2020orz, Heidenreich:2021tna}.
One example of this is the D7-brane of Type IIB string theory, which induces a monodromy of the Type IIB axion $C_0$ upon winding around the core of the D7-brane.

As in the case of the abelian Higgs model, the decay parameter $f$ of a fundamental axion is controlled by the vev of a ``saxion'' scalar field:
\begin{equation}
S = - \frac{1}{2}  \int d^dx \sqrt{-\mathfrak{g}_d}   \left[  g(f) (\partial_\mu f)^2 + f^2 (\partial_\mu \theta)^2  \right]\,,
\label{action}
\end{equation}
where we have assumed that the metric on moduli space $g_{ff} = g(f)$ does not depend on $\theta$, and we may further assume that the spacetime metric $\mathfrak{g}_d$ is flat.\footnote{Adding an Einstein-Hilbert term $S_{EH} =- \frac{1}{2}  \int d^dx \sqrt{-\mathfrak{g}_d}  \mathcal R_d$ and promoting gravity to a dynamical field, the Einstein equations are solved by the metric $ds_d^2 = \eta^{(d-2)}_{\mu\nu}dx^\mu dx^\nu + e^{2\Omega(r)}(dr^2 + r^2 d \varphi^2)$ for a suitable choice of the warp factor $\Omega(r)$. The inclusion of this warp factor has no effect on our results in this section, as we explain further in an appendix.}

By requiring that the gradient energy of the vortex solution remains finite as $\theta$ winds around its fundamental domain, we see that the parameter $f$ must vanish at the core of the string, ensuring that Conjecture 4 is satisfied. Thus, we see that Conjecture 4 is intimately connected to the presence of axion vortex solutions, which is in turn related to the absence of $(d-2)$-form symmetries in quantum gravity.

In fact, the requirement of finite-energy vortex solutions imposes even tighter constraints on the form of the effective action \eqref{action}.  Assuming a spherically symmetric ansatz $f=f(r)$, $\theta = \varphi$ near the core of the vortex,\footnote{Note that in the example of the D7-brane, this spherical symmetry eventually breaks down some distance from the core of the string, but it is valid near the core \cite{Greene:1989ya, Gibbons:1995vg}.} where $(r, \varphi)$ are the polar coordinates of the transverse plane, the requirements of (a) infinite-distance and (b) finite energy imply $h(r) \equiv \sqrt{g(f(r))} f'(r) \sim r^{-1} |\log r/r_0|^{-\gamma}$,
where $\frac{1}{2} < \gamma \leq 1$. Further imposing the equations of motion implies $f(r) = r h(r)$ up to a constant of integration.
In terms of the proper field distance $\Delta$, this implies that the decay constant $f$ vanishes in the limit $r \rightarrow 0$ as
\begin{equation}
f \sim \left\{ \begin{array}{cc}
\Delta^{ \frac{-\gamma}{1-\gamma}} & \frac{1}{2} < \gamma < 1  \\
\exp( -\alpha \Delta) & \gamma =1
\end{array}
 \right.,
 \label{fscaling}
\end{equation} 
for some constant $\alpha$.\footnote{The $\gamma=1$ case is common in string theory---for instance, the D7-brane in Type IIB string theory has $\gamma = 1$, $\alpha = \sqrt{2}$.}
In other words, $f$ must vanish more quickly than $1/\Delta$, and it can vanish no more quickly than an exponential in the proper field distance $\Delta$.

The upshot of this is that the effective action for the saxion-axion system is constrained by the requirement that dynamical vortex solutions exist; these vortices are in turn required in order to break the $(d-2)$-form symmetry associated with a nontrivial $\pi_1(\cM)$. The fact that $f$ vanishes near the core of the string ensures that Conjecture 4 is satisfied, and it often plays an important role in satisfying Conjecture 3, as we now explain.

\section*{Conjecture 3 and axion vortices}

 \vspace{.2cm}
\begin{namedconjecture}[Conjecture 3]
The scalar curvature near the points at infinity is non-positive. (It is
strictly negative if the dimension of the moduli space is greater than 1.)
\end{namedconjecture}
 \vspace{.2cm}

\noindent
Conjecture 3 is the one Ooguri-Vafa conjecture which has been falsified: in \cite{trenner2010asymptotic}, Trenner and Wilson found examples of Calabi-Yau manifolds whose scalar curvature is positive in certain asymptotic limits. More recently, it was argued that the positive curvature in these examples is related to the existence of a light sector that decouples from gravity in the infinite-distance limit \cite{Marchesano:2023thx}. Nonetheless, Conjecture 3 is valid in many examples of moduli spaces in string theory where such a decoupled sector does not exist, and thus it points us toward common, if not entirely universal, features of quantum gravity.

Infinite-distance limits in quantum gravity typically feature a flat slice of moduli space parametrized by one or more noncompact scalar fields. In addition, they have a number of compact scalar fields, or axions, whose decay parameters depend on the noncompact scalar fields. These compact directions in moduli space are responsible for the negative curvature of the moduli space. In the simplest case of a single axion coupled to a noncompact saxion scalar field, this negative curvature can be seen from the analysis above.

In particular, consider once again the action \eqref{action}. In terms of the canonically normalized saxion field $\rho$, the
metric on moduli space is given by $g_{ij} = \text{diag}(1,f(\rho)^2)$, and the scalar curvature for this metric is given by $R = - 2{f''(\rho)}/{f}$.
In the infinite-distance limit at the core of the axion vortex, we have $\rho \rightarrow \infty$, $f \rightarrow 0$, $f > 0$. Setting $\Delta = \rho$ in \eqref{fscaling}, we see that $f''(\rho) > 0$ in this limit, so indeed $R<0$.

Thus, at least in this toy model, the presence of axion strings is closely related to the negative curvature of the moduli space in the asymptotic limit. More generally, it has been conjectured that every infinite-distance limit of a 4d effective field theory is realized at the core of some axion string \cite{Lanza:2021qsu}. This offers a possible explanation as to why negative curvature is so prevalent in quantum gravity moduli spaces, though again we stress that it is not quite universal.

Finally, let us remark that---as discussed in \cite{Ooguri:2006in}---the negative curvature in asymptotic limits of moduli space is closely related to the finiteness of their volume.\footnote{In \cite{Vafa:2005ui}, Vafa noted that moduli spaces in quantum gravity of dimension greater than one seem to have finite volume.}
In the example above, the volume of moduli space in the infinite-distance limit is given by the integral
$\text{vol}(\cM) = \int d \omega = \int d \rho d \theta f(\rho)$,
where $\omega$ is the volume form on moduli space. Plugging in the scaling of $f(\rho)$ from \eqref{fscaling}, we see that the lower bound $\gamma > 1/2$ is precisely the condition that the volume of the $\rho$-$\theta$ moduli space remains finite in the limit $\rho \rightarrow \infty$.

\section*{Conjecture 1 and Dynamical Cobordisms}

 \vspace{.2cm}
\begin{namedconjecture}[Conjecture 1]
Choose any point $p_0 \in \cM$. For any positive $T$, there is another point $p \in \cM$ such that $||p-p_0|| > T$.
\end{namedconjecture}
 \vspace{.2cm}

\noindent
In other words, Conjecture 1 holds that the diameter of moduli space is infinite. 

In the axion vortex example discussed above, the infinite diameter of moduli space was related to the presence of dynamical axion vortices. These vortices are responsible for breaking a (possibly non-invertible) higher-form symmetry. In modern language, they ensure that the codimension-2 operators charged under the would-be symmetry can end on operators of codimension-3, which implies that the winding number around these codimension-2 operators is no longer a valid conserved charge.

This represents just one example of a more general lesson: breaking a higher-form global symmetries can be understood as the statement that certain extended operators of dimension $p$ can end on operators of dimension $p-1$. Mathematically speaking, this ``endability'' is related to the notion of cobordism. Two manifolds $M$, $N$ of dimension $d$ are said to be cobordant if there exists a $(d+1)$-dimensional manifold whose boundary is the disjoint union $M \sqcup N$. This defines an equivalence relation between the manifolds $M$ and $N$, and the equivalence classes can be imbued with a group structure, with disjoint union as the group multiplication law.

Within quantum field theory, the statement that an extended operator can end on a lower-dimension operator is tantamount to the statement that any operator which links it must be trivial in cobordism, and it in turn implies that that the endable operator cannot be charged under a global symmetry. As a result, the statement that there are no global higher-form symmetries in quantum gravity is more or less equivalent to the statement that all cobordism groups in quantum gravity are trivial \cite{McNamara:2019rup}.

With this, the axion vortex solutions discussed above are just one example of a more general class of ``dynamical cobordism defects,'' \cite{Buratti:2021fiv, Blumenhagen:2022mqw, Angius:2022aeq, Angius:2022mgh, Blumenhagen:2023abk} which (a) trivialize cobordism groups by allowing certain extended operators to end, and (b) involve a scalar field which runs to infinite distance at their core. As explained in \cite{Angius:2022aeq}, D-branes in string theory can be understood as examples of such dynamical cobordism defects.

These cobordism defects suggest a close relationship between infinite-distance limits in scalar field space and the absence of nontrivial cobordism groups. Making this relationship precise is tricky because there are also examples of cobordism defects in string theory that do not involve infinite-distance limits in field space---most famously, the Horava-Witten wall can be understood as a cobordism defect of M-theory \cite{McNamara:2019rup}, even though M-theory has no massless moduli at all. By clarifying the conditions under which infinite-distance limits are required for the existence of cobordism defects, one may be able to put Conjecture 1 on firmer footing.

\section*{Conjecture 2 and approximate global symmetries}

 \vspace{.2cm}
\begin{namedconjecture}[Conjecture 2]
Compared to the theory at $p_0 \in \cM$, the theory at $p$ with $||p-p_0|| > T$ has
an infinite tower of light particles starting with mass of the order of $\e^{- \alpha T}$ for some $\alpha > 0$
In the $T \rightarrow \infty$ limit, the number of extra light particles of mass less than a fixed mass
scale becomes infinite.
\end{namedconjecture}
 \vspace{.2cm}
 
\noindent
Also known as the ``Distance Conjecture,'' this conjecture may be related to the breaking of approximate global symmetries in quantum gravity. 

A free, massless scalar field $\phi$ has an exact shift symmetry. Coupling this scalar field to a fermion $\psi$ with a Yukawa-like coupling $\lambda m_0(\phi) \bar \psi \psi$, this symmetry is restored in the limit $\lambda \rightarrow 0$. Thus, for small $\lambda$, we have an approximate global symmetry. 

Naively, this suggests that a weakly coupled scalar field theory necessarily leads to an approximate symmetry. However, this approximate symmetry will be badly broken provided there are a sufficient number of light fields coupled weakly to the scalar, as these run in loops and drive the scalar field toward strong coupling in the UV. As shown in \cite{Cordova:2022rer} (based on earlier calculations in \cite{Grimm:2018ohb, Heidenreich:2018kpg, Heidenreich:2017sim}), the requirement that the symmetry is badly broken within the low-energy EFT is equivalent (up to order-one factors) to the statement that the fermion masses $\langle m_0(\phi) \rangle$ decay exponentially with proper field distance, . 

At present, this connection between Conjecture 2 and the breaking of approximate global symmetries is suggestive but rough: there is no precise argument against approximate global symmetries in quantum gravity, and it is not clear how to extend the argument of \cite{Cordova:2022rer} to towers of string oscillation modes, which also decay exponentially in infinite-distance limits but are not well-described by low-energy EFT. Making these arguments more precise would be an important step toward a proof of Conjecture 2.

.

\section*{Conclusions}

In this essay, we have seen that generalized notions of global symmetries are closely related to all of the Ooguri-Vafa conjectures. The requirement that these symmetries must be gauged or broken motivates many observed features of scalar field theories in quantum gravity, and it helps to shed light on these theories beyond the supersymmetric lamppost of string theory.

Yet at the same time, many open questions remain. What limitations does quantum gravity place on approximate global symmetries, and why? What can statements about asymptotic regimes in scalar field spaces tell us about our own universe, which likely lives somewhere in the interior? Can holography, quantum information theory, and pure mathematics help shed further light on scalar fields in quantum gravity?  At the heart of these questions lie answers to some of the most fundamental mysteries of our universe.

\section*{Acknowledgements}

It is a pleasure to thank I\~naki Garc\'ia-Etxebarria, Ben Heidenreich, Jacob McNamara, Matthew Reece, Mykhaylo Usatyuk, and Timo Weigand for useful discussions. This work was supported in part
by STFC through grant ST/T000708/1.

\appendix

\section{Axion vortices}

In this appendix, we include additional details of the calculations on axion vortices in the section on Conjecture 4. We begin with the action of \eqref{action}, plus an additional Einstein-Hilbert term:
\begin{equation}
S = - \frac{1}{2}  \int d^dx \sqrt{-\mathfrak{g}_d}   \left[ - \mathcal R_d +  g(f) (\partial_\mu f)^2 + f^2 (\partial_\mu \theta)^2  \right]\,.
\label{actionappen}
\end{equation}
Once again, we have assumed that the metric on moduli space $g_{ff} = g(f)$ does not depend on $\theta$.
We take a metric ansatz of the form,
\begin{equation}
d s_d^2 = ds_{d-2}^2 + e^{2\Omega(r)}(dr^2 + r^2 d \varphi^2),
\end{equation}
where $\Omega(r)$ is the warp factor and $ds_{d-2}^2 = \eta_{\mu\nu}^{(d-2)} dx^\mu dx^\nu$ is taken to be flat, $(d-2)$-dimensional Minkowski space. With this, the action \eqref{actionappen} becomes
\begin{equation}
S = - \frac{1}{2}  \int d^{d-2}x dr d\varphi   r e^{2\Omega(r)}   \left[-\mathcal R_{d} + g(f) (\partial_\mu f)^2 + f^2 (\partial_\mu \theta)^2  \right]\,,
\end{equation}

We next take a spherically symmetric ansatz near the core of the string of the form
\begin{equation}
f=f(r)\,,~~~\theta = \varphi\,,
\label{ansa}
\end{equation}
where $(r, \varphi)$ are the polar coordinates of the transverse plane.
We then insist that there exists a vortex solution with nontrivial winding and finite energy per unit length.
The energy per unit length from the gradient of the scalar fields is given by
\begin{align}
\mathcal{E} =  \frac{1}{2} \int r dr d \varphi \left[ g(f) f'(r)^2 + \frac{f(r)^2}{(2 \pi r)^2}    \right] \,.
\label{Eeq}
\end{align}
Notably, the warp factor does not appear in this expression, as the factor of $\exp(2 \Omega(r))$ from the square root of the determinant has canceled the factor of $\exp(-2 \Omega(r))$ from the inverse metric in the kinetic terms. The requirement that $\mathcal{E} $ is finite implies that both terms in the integrand must integrate to a finite value. However, we have also assumed that the limit $f \rightarrow 0$ is at infinite distance in field space, which means that 
\begin{align}
||\Delta f|| = \int dr \sqrt{g(f)} f'(r) 
\label{fsdist}
\end{align}
must diverge. Together, these constraints are not easy to satisfy! Setting $h(r) = \sqrt{g(f(r))} f'(r)$, we have
\begin{align}
\int_0^{r_0} dr h(r) \rightarrow \infty \,,~~~\int_0^{r_0} r dr  h(r)^2 = \text{finite}\,.
\end{align}
It is easy to see that $h(r)$ cannot satisfy these constraints if it simply scales like a power law near $r=0$; instead, we must have 
\begin{align}
h(r) \sim \frac{1}{r |\log r/r_0|^\gamma}\,,
\label{finiteE}
\end{align}
where $\frac{1}{2} < \gamma \leq 1$.

This already fixes the scaling behavior of the canonically-normalized saxion field, but we can go even further by imposing the equation of motions for the fields $f$ and $\theta$:
\begin{align}
\partial^{\mu} (r g \partial_\mu f) = \frac{1}{2}  r g'(f) (\partial_\mu f)^2 + f  r (\partial_\mu \theta)^2 
\label{feom}\\
\partial^\mu r f^2 \partial_\mu \theta = 0\,.
\label{thetaeom}
\end{align}
Once again, powers of the warp factor $\exp(\Omega(r))$ have canceled out in these equations.
Assuming the spherically symmetric ansatz \eqref{ansa}, the $\theta$ equation of motion is satisfied trivially, while the equation of motion for $f$ becomes
\begin{equation}
g'(f) f'(r)^2 + g f''(r) + \frac{1}{r} g f'(r) = \frac{1}{2} g'(f) f'(r)^2 + \frac{f}{r^2}\,.
\end{equation}
Plugging in $h(r) = \sqrt{g} f'(r)$, we can massage this equation into the simple form
\begin{equation}
\frac{\partial}{\partial r} (h^2 r^2) = \frac{\partial}{\partial r} (f^2)\,.
\end{equation} 
So, 
\begin{equation}
 h^2 r^2 = f^2 \,,
 \label{hfeq}
\end{equation} 
up to a constant of integration. Using the form of $h(r)$ in \eqref{finiteE}, we conclude that
\begin{equation}
f(r) \sim |\log r/r_0 |^{-\gamma}\,,
\end{equation}
and therefore
\begin{equation}
g(f(r)) = \frac{h^2}{(f'(r))^2} \sim |\log r / r_0|^2 ~~~ \Rightarrow ~~~ g(f) \sim f^{-2/\gamma}\,.
\end{equation}

%

Thus, the requirement of a spherically-symmetric, finite-energy vortex solution to the equations of motion near $r=0$ implies that the effective action takes the form
\begin{equation}
S = - \frac{1}{2}  \int d^dx \sqrt{-\mathfrak{g}_d}   \left[ \frac{g_0}{f^{2/\gamma}} (\partial_\mu f)^2 + f^2 (\partial_\mu \theta)^2  \right]\,,
\end{equation} 
where $\frac{1}{2} < \gamma \leq 1$.\footnote{This same range of possibilities was also found in \cite{Lanza:2021qsu} in the specific context of BPS strings in 4d supergravity theories.}

This form of the saxion-axion action is ubiquitous in string theory. For instance, the Type IIB action takes this form \cite{Polchinski:1998rr}, with $\gamma=1$, $\theta = 2 \pi C_0$, $f = (2 \sqrt{2} \pi \tau_2)^{-1}$, and $g_0 =1/2$. In this context, the particular choice $g_0= 1/2$ is required by $SL(2, \mathbb{Z})$ duality.

Finally, let us introduce a canonically normalized saxion field $\rho$ via $ d \rho = - \sqrt{g_0} f^{-1/\gamma} df$. With this, the limit $f \rightarrow 0$ corresponds to $\rho \rightarrow \infty$. In terms of the proper field distance $\Delta = \rho$, the decay constant $f$ therefore vanishes as
\begin{equation}
f \sim \left\{ \begin{array}{cc}
\Delta^{ \frac{-\gamma}{1-\gamma}} & \frac{1}{2} < \gamma < 1  \\
\exp( -\alpha \Delta) & \gamma =1
\end{array}
 \right.,
 \label{fscaling}
\end{equation}
where $ \alpha = \frac{1}{\sqrt{g_0}}$.
From the limiting cases $\gamma =1/2$ and $\gamma=1$, we find an lower bound and an upper bound on the rate at which the decay constant $f$ may tend to zero with proper field distance, namely
\begin{equation}
f \Delta \rightarrow 0  \text{ as } {\Delta \rightarrow \infty} \,,~~~~ \exp(- \alpha \Delta) \lesssim f  \,.
\end{equation}
In other words, $f$ must vanish more quickly than $1/\Delta$, and it can vanish no more quickly than an exponential in the proper field distance $\Delta$.

%

Note that the inclusion of gravity had no effect on the calculations above, as the warp factor did not appear in the gradient energy \eqref{Eeq}, the field space distance \eqref{fsdist}, or the equations of motion \eqref{feom}-\eqref{thetaeom}.
The only significant effect of including gravity comes from the Einstein field equation $G_{tt} = T_{tt}$, which fixes the form of the warp factor $\Omega(r)$ near the core of the vortex $r=0$:
\begin{equation}
-e^{-2 \Omega}\frac{1}{r}\partial_r(r \Omega'(r)) = e^{-2 \Omega} \left[\frac{1}{2}g(f) f'(r)^2 + \frac{f^2}{ 2r^2} (\partial_\varphi\theta)^2\right] \,.
\end{equation}
Setting $\theta = \varphi$, $h(r) = \sqrt{g}f'(r)$, and using the result $f(r) = r h(r)$ from \eqref{hfeq}, this becomes
\begin{equation}
    \frac{1}{r}\partial_r(r \Omega'(r)) = -h(r)^2\,.
\end{equation}
Setting $h(r) \sim \frac{1}{r |\log r/r_0|^\gamma}$ near $r=0$, we find that this equation is solved by
\begin{equation}
    \Omega(r) \sim \left\{ \begin{array}{cc}
|\log r/r_0|^{2 - 2 \gamma} & \frac{1}{2} < \gamma < 1  \\
|\log|\log r/r_0|| & \gamma =1
\end{array}
 \right..
\end{equation}
We see that the warp factor $\Omega(r)$ tends to $+\infty$ as $r \rightarrow 0$, but $\int_0^\epsilon dr \exp(\Omega(r))$ converges for small $\epsilon > 0$, so the spacetime distance to the vortex core remains finite.

\bibliographystyle{utphys}
\bibliography{ref}
\end{document}